\documentstyle[12pt]{article}
\def\MeV{\,{\rm MeV}}

\def\cmm2{{\,\rm cm^{-2}}}
\def\cm2{{\,{\rm cm}^2}}
\def\cmm3{{\,{\rm cm}^{-3}}}
\def\gcmm3{{\,{\rm g\,cm^{-3}}}}

\def\la{\mathrel{\mathpalette\fun <}}
\def\ga{\mathrel{\mathpalette\fun >}}
\def\fun#1#2{\lower3.6pt\vbox{\baselineskip0pt\lineskip.9pt
  \ialign{$\mathsurround=0pt#1\hfil##\hfil$\crcr#2\crcr\sim\crcr}}}

\makeatletter

\let\old@bibitem=\@bibitem
\def\@bibitem#1#2{\@ifundefined{r@#1}{}{\@warning
  {Multiple entries for reference `#1'}}\@ifundefined{b@#1}{\@warning
  {Reference `#1' not cited}}{\global\@namedef{r@#1}{#2}}}
\let\old@lbibitem=\@lbibitem
\def\@lbibitem[#1]#2#3{\@ifundefined{r@#2}{}{\@warning
  {Multiple entries for reference `#2'}}\@ifundefined{b@#2}{\@warning
  {Reference `#2' not cited}}{\global\@namedef{r@#2}{#3}
  \global\@namedef{s@#2}{#1}}}

\def\@putbibrefs#1{\expandafter\@makearef #1*,}
\def\@makearef#1,{\if*#1 \let\next=\relax \else \let\next=\@makearef
  \@ifundefined{s@#1}{\bibitem{#1}}{\bibitem[\@nameuse{s@#1}]{#1}}
  \@ifundefined{r@#1}{Citation label `#1'?}{\@nameuse{r@#1}}
  \fi \next}

\def\endthebibliography{\let\@bibitem=\old@bibitem \let\@lbibitem=\old@lbibitem
  \@putbibrefs\@bibrefslist \endlist}

\def\@citex[#1]#2{\if@filesw\immediate\write\@auxout{\string\citation{#2}}\fi
  \def\@citea{}\@cite{\@for\@citeb:=#2\do
    {\@citea\def\@citea{,\penalty\@m\ }\@ifundefined
       {b@\@citeb}{{\bf ?}\@warning
       {Citation `\@citeb' on page \thepage \space undefined}}%
\hbox{\csname b@\@citeb\endcsname}\@ifundefined
       {d@\@citeb}{\global\@namedef{d@\@citeb}{y}\@addlabel{\@citeb}}{}}}{#1}}

\def\nocite#1{\@ifundefined{d@#1}{\@addlabel{#1}}{\@warning
	{Label `#1' cited.  \string\nocite\space not necessary.}}}

\def\@addlabel#1{\@ifundefined{@bibrefslist}{\xdef\@bibrefslist{#1,}}{\xdef
	\@bibrefslist{\@bibrefslist#1,}}}
\makeatother

\begin{document}
\pagestyle{empty}
\begin{center}
\bigskip

\rightline{FERMILAB--Pub--95/257-A}
\rightline{astro-ph/9508029}
\rightline{submitted to {\it Physical Review Letters}}
\vspace{.2in}
{\Large \bf Assessing Big-Bang Nucleosynthesis}
\bigskip

\vspace{.2in}
Craig J.~Copi,$^{1,2}$ David N.~Schramm,$^{1,2,3}$ and
Michael S. Turner$^{1,2,3}$\\

\vspace{.2in}
{\it $^1$Department of Physics \\
Enrico Fermi Institute, The University of Chicago, Chicago, IL~~60637-1433}\\

\vspace{0.1in}
{\it $^2$NASA/Fermilab Astrophysics Center\\
Fermi National Accelerator Laboratory, Batavia, IL~~60510-0500}\\

\vspace{0.1in}
{\it $^3$Department of Astronomy \& Astrophysics\\
The University of Chicago, Chicago, IL~~60637-1433}\\

\end{center}

\vspace{.3in}
\centerline{\bf ABSTRACT}
\bigskip

Systematic uncertainties in the light-element abundances and their evolution
make a rigorous statistical assessment difficult.  However, using
Bayesian methods we show that the following statement is robust:
the predicted and measured abundances are consistent with 95\% credibility
only if the baryon-to-photon ratio is between
$2\times 10^{-10}$ and $6.5\times 10^{-10}$ and the number
of light neutrino species is less than 3.9.  Our analysis
suggests that the $^4$He abundance may have been systematically underestimated.


\newpage
\pagestyle{plain}
\setcounter{page}{1}
\newpage

Big-bang nucleosynthesis occurred seconds after the bang
and for this reason offers the most stringent test of the
standard cosmology.  Comparison of the predicted and
measured light-element abundances has evolved dramatically
over the past thirty years, beginning with the observation
that there was evidence for a significant primeval abundance
of $^4$He which could be explained by the big bang \cite{hoyle}
to the present where the abundances of D, $^3$He, $^4$He and
$^7$Li are all used to test the big bang.

The predictions of big-bang nucleosynthesis depend upon the
baryon-to-photon ratio ($\equiv \eta$) as well as the number
of light ($\la 1 \MeV$) particle species, often quantified as
the equivalent number of massless neutrino species ($\equiv N_\nu$).
For a decade it has been argued that the abundances of
all four light elements can be accounted for provided $\eta$ is
between $2.5\times 10^{-10}$ and $6\times 10^{-10}$ and
$N_\nu < 3.1-4$ \cite{cst,Nnu,yangetal}.
The ``consistency interval''  provides
the best determination of the baryon density and is key to the case for
the existence of nonbaryonic dark matter.  The limit to $N_\nu$
provides a crucial hurdle for theories that aspire to
unify the fundamental forces and particles.

However, these conclusions were not based upon a rigorous statistical analysis.
Because the dominant uncertainties in the light-element
abundances are systematic such an analysis is difficult
and previous work focussed on concordance intervals.
Given the importance of big-bang nucleosynthesis
it is worthwhile to try to use more rigorous
methods.  Here we apply two standard methods, goodness of fit
and Bayesian likelihood, and identify the
the conclusions which are insensitive to the systematic errors.

We begin by reviewing the general situation.  The predictions of standard
big-bang nucleosynthesis are shown in Fig.~1.  The theoretical uncertainties
are statistical, arising from imprecise knowledge of the neutron lifetime and
certain nuclear cross sections.  Because of 10 Gyr or so of ``chemical
evolution'' since the big bang (nuclear reactions in stars and elsewhere which
modify the light-element abundances) determining primeval abundances is not
simple and the dominant uncertainties are systematic.

The chemical evolution of $^4$He is straightforward:
stars make additional $^4$He.  Stars also make metals
(elements heavier than $^4$He); by
measuring the $^4$He abundance in metal-poor,
extragalactic HII (ionized hydrogen) clouds as a function of
some metal indicator
(e.g., C, N or O) and extrapolating to zero metallicity
a primeval abundance has been inferred:  $Y_P =0.232\pm 0.003\,{\rm (stat)}
\pm 0.005\,{\rm (sys)}$ \cite{os}.  Systematic uncertainties arise from trying
to convert line strengths to abundances by modeling.
The range $Y_P=0.221 - 0.243$ allows for $2\sigma$
statistical + $1\sigma$ systematic uncertainty and
is consistent with the big-bang prediction provided $\eta
\simeq (0.8 - 4) \times 10^{-10}$ \cite{cst}.
Others have argued that the systematic uncertainty is a factor of two or
even three larger \cite{he4sys}; taking $Y_P \simeq
0.21 - 0.25$ increases the concordance range significantly,
$\eta \simeq (0.6-10) \times 10^{-10}$, reflecting the logarithmic dependence
of big-bang $^4$He production upon $\eta$ \cite{cst}.

There is a strong case that the $^7$Li abundance measured in
metal-poor, old pop II halo stars, $^7{\rm Li/H}
= (1.5\pm 0.3)\times 10^{-10}$, reflects the big-bang abundance \cite{li7}.
However, it is possible that even in these stars the $^7$Li
abundance has been reduced by nuclear burning, perhaps by a
factor of two (the presence of $^6$Li in some of these stars,
which is more fragile, provides an upper limit to the amount of astration).
Allowing for a $2\sigma$ statistical uncertainty
and for up to a factor of two astration,\footnote{Additional smaller,
but important, systematic uncertainties arise from the modeling of stellar
atmospheres, which is needed to convert line strengths to
abundances, and possible enhancement of $^7$Li by cosmic-ray
production \cite{cst}.   We use astration
to illustrate the effects of systematic uncertainty in the
$^7$Li abundance, but it serves to illustrate the point for the other
effects too.}  leads to the consistency interval
$\eta = (1- 6)\times 10^{-10}$ \cite{cst}.

The interpretation of D is particularly challenging because it is burned in
virtually all astrophysical situations and its abundance has only been
accurately measured in the solar vicinity.  Because D is destroyed and not
produced \cite{els} and because its abundance is so
sensitive to $\eta$ (D/H $\propto \eta^{-1.7}$), a firm upper limit to $\eta$
can be obtained by insisting that big-bang production account for the D
observed locally, D/H $\ga (1.6\pm 0.1) \times 10^{-5}$ \cite{hst}.
This leads to the two-decade old bound, $\eta\la 9\times 10^{-10}$, which is
the linchpin in the argument that baryons cannot provide closure density
\cite{reevesetal}.  Because D is readily destroyed,
it is not possible to use the D abundance
to obtain a lower bound to $\eta$.  The sum of D and $^3$He
is more promising:  D is first burned to $^3$He,
and $^3$He is much more difficult to burn.  On the
assumption that the $^3$He survival fraction is greater than 25\%
the lower limit $\eta \ga 2.5\times 10^{-10}$
has been derived \cite{yangetal}.  The D, $^3$He concordance
interval, $\eta \simeq (2.5-9) \times 10^{-10}$.

The overlap of the concordance intervals (see Fig.~1),
which occurs for $\eta = (2.5- 6) \times 10^{-10}$,
is the basis for concluding that the light-element
abundances are consistent with the big-bang predictions \cite{cst}.

The dominant uncertainties in comparing the predicted and
measured light-element abundances are
systematic:  the primeval abundance of $^4$He;
the chemical evolution of D and $^3$He; and whether or not $^7$Li
in the oldest stars has been reduced significantly by nuclear
burning.  Systematic error is difficult to treat as it is
usually poorly understood and poorly quantified.  (If it were understood
and were well quantified it wouldn't be systematic error!)
This is especially true for astronomical {\it observations},
where the observer has little control over the object being observed.

There are at least three kinds of systematic error.
(1) A definitive, but unknown, offset between what is measured and
what is of interest.  (2) A random source of error whose distribution is
poorly known.  (3) An important source of error that is unknown.
The first kind of systematic error is best treated
as an additional parameter in the likelihood function.
The second kind of systematic error is best
treated by use of a distribution, or by several candidate
distributions.  The third type of systematic error is a nightmare.

The data themselves can clarify
matters.  Consider $^7$Li; its measured abundance
in old pop II stars is equal to the primeval abundance with
a small statistical error and a larger
systematic uncertainty due to astration.  This could be a systematic
error of the first kind---if all stars reduce their $^7$Li
abundance by the same factor---or of the second kind---if the $^7$Li
abundance in different stars were reduced by different amounts.
In the latter case, the measured $^7$Li abundance should show a
large dispersion---which it does not \cite{li7}.  Thus, we treat
astration by considering two limiting possibilities:
$^7$Li/H$ = (1.5\pm 0.3)\times 10^{-10}$ (no astration);
and second, $^7$Li/H$ = (3.0\pm 0.6)\times 10^{-10}$
(astration by a factor of two).

Several sources of systematic error for $^4$He have been
identified---dust absorption, neutral $^4$He, stellar
absorption, and theoretical emissivities---which can reduce
or increase the measured abundance \cite{he4sys}.  If the same effect
dominates in each measurement use of an offset parameter in the
$^4$He abundance would be
appropriate.  On the other hand, if different effects dominate
different measurements enlarging the statistical
error would be appropriate.  We allow for both:  the statistical
error $\sigma_Y$ is permitted to be larger than 0.003, and an offset
in the $^4$He abundance, $\Delta Y$, is a parameter
in the likelihood function ($Y_P = 0.232 +\Delta Y$).

Finally, there is the systematic uncertainty associated with the
chemical evolution of D and $^3$He.  Based
upon a recent study of the chemical evolution of D and $^3$He \cite{cst2} we
consider three models that encompass the broadest
range of possibilities:  Model 0 is the plain, vanilla model; Model 1 is
characterized by extreme $^3$He destruction\footnote{Because the
stars that destroy $^3$He also make metals, it is not possible
to destroy $^3$He to an arbitrary degree without overproducing
metals \cite{cst2}.}
(average $^3$He survival factor of about 15\%); and Model 2 is
characterized by minimal $^3$He destruction.  The
likelihood functions for these three models are shown in Fig.~2.

First consider the $\chi^2$ test for goodness of fit.
This technique is best suited to situations where the errors
are gaussian and well understood and there are many
degrees of freedom; neither apply here.  Nonetheless, in
Fig.~3 we show $\chi^2 (\eta )$ for
eight different assumptions about the systematic uncertainties:
(1,5) $\sigma_Y = 0.003$, $\Delta Y = 0$;
(2,6) $\sigma_Y = 0.01$, $\Delta Y = 0$;
(3,7) $\sigma_Y = 0.003$, $\Delta Y = 0.01$; (4,8) $\sigma_Y =
0.01$, $\Delta Y = 0.01$.  In (1)--(4), $^7{\rm Li/H}=(3\pm 0.6)\times
10^{-10}$; and in (5)--(8) $^7{\rm Li/H}=(1.5\pm 0.3)\times
10^{-10}$.  For clarity, only the results for Model 0 are shown,
the results for Models 1 and 2 are similar.

Several conclusions can be drawn from Fig.~3.  First, the goodness of fit
depends sensitively upon assumptions made about the systematic errors, with
the minimum $\chi^2$ ranging from 6 to much less than 1; it is smallest when a
systematic shift in $^4$He is allowed and/or $\sigma_Y$ is increased.  Second,
in all cases the the $\eta$ interval defined by $\Delta\chi^2 =3$ (from the
minimum $\chi^2$) has a lower bound no lower than about $1.5\times 10^{-10}$
(in all but (5), no lower than $2.5\times 10^{-10}$) and an upper bound no
higher than about $6\times 10^{-10}$.

Next, we turn to Bayesian likelihood, which is best suited
to determining parameters of a theory or assessing the relative viability
of two or more theories.  Since there is no well developed,
alternative to the standard theory of
nucleosynthesis at present, likelihood is of no use
in assessing relative viability.  Systematic errors of the first
kind are treated as additional (nuisance) parameters
in the likelihood function which can be determined
by the experiment itself or can be eliminated by marginalization;
we treat $\Delta Y$ as such.  We also
allow $\sigma_Y$ to vary to study how results depend upon the
assumed uncertainty in the $^4$He abundance.  Because we are interested
in setting a limit to $N_\nu$, it too is taken to be a parameter.
Values of $N_\nu$ greater than three describe extensions of the
standard model with additional light degrees of freedom.

In Fig.~4 we show contours of
${\cal L}(N_\nu,\Delta Y, \sigma_Y =0.003)$.  The contours
are diagonal lines because $\Delta Y$
and $N_\nu$ are not independent parameters---the primary effect
of an increase in $N_\nu$ is an increase in the predicted
$^4$He abundance ($\Delta Y_P \sim 0.01\,\Delta N_\nu $).
A likelihood function that is not compact must be treated with care,
because no information
about the parameters (here, $N_\nu$ and $\Delta Y$) can be inferred
independently of what was already known (the priors).

For example, the likelihood function ${\cal L}(N_\nu)$, which is needed to set
limits to $N_\nu$, is obtained by integrating over $\Delta Y$ and depends upon
the limits of integration.  To derive limits to $N_\nu$ we do the following:
integrate from $-|\Delta Y|$ to $|\Delta Y|$; normalize ${\cal L}(N_\nu)$ to
have unit likelihood from $N_\nu =3$ to $\infty$; the limit is the value of
$N_\nu$ beyond which 5\% of the total likelihood accumulates.  The dependence
of the limit upon $|\Delta Y|$ is shown in Table~1.

\begin{table}
\begin{center}
\begin{tabular}{cccc} \hline
$|\Delta Y|$ & Model 0 & Model 1 & Model 2 \\ \hline
0 & 3.1/3.1 & 3.2/3.3 & 3.1/3.1 \\
0.005 & 3.2/3.1 & 3.3/3.3 & 3.2/3.1 \\
0.010 & 3.3/3.2 & 3.5/3.3 & 3.3/3.2 \\
0.015 & 3.5/3.4 & 3.7/3.4 & 3.4/3.4 \\
0.020 & 3.7/3.6 & 3.9/3.7 & 3.7/3.6 \\ \hline
\end{tabular}
\end{center}
\caption{Limits to $N_\nu$ for Models 0, 1, 2 and
${\rm Li/H}=(1.5 \pm 0.3) \times
10^{-10}$ (first number) and ${\rm Li/H}=(3.0\pm0.6)\times 10^{-10}$
(second number).}
\end{table}

An aside; in a recent paper the likelihood function
${\cal L}(N_\nu)$ obtained by integrating from $\Delta Y = -0.005$
to $0.005$ was used in an attempt to assess the viability of
the standard theory of nucleosynthesis \cite{hata}.
This likelihood function is peaked at
$N_\nu = 2$ and is approximately gaussian with $\sigma_{N_\nu}
=0.3$.  On this basis it was claimed that the standard theory of
nucleosynthesis is ruled out with 99.7\% confidence.
By so doing equal weight was implicitly given
to all values of $N_\nu$ (uniform priors).
The prior for $N_\nu=3$
(the standard model of particle physics) is certainly orders
of magnitude greater than that for $N_\nu =2$ (for which no
well developed model exists).   The likelihood function
${\cal L}(N_\nu)$ which properly included prior information would
certainly not be peaked at $N_\nu = 2.0$.

In Figs. 5 and 6 we show the 95\% contours of the
likelihood functions ${\cal L}(\eta, \sigma_Y )$ and
${\cal L}(\eta, \Delta Y)$ for Models 0, 1, and 2
and both values of the central $^7$Li abundance.
Both figures suggest the same thing:  the uncertainty in the
primordial $^4$He abundance has been underestimated.  In the
$\sigma_Y-\eta$ plane $\sigma_Y = 0.003$ does not intersect
the 95\% credibility contour, and in the $\Delta Y - \eta$ plane
$\Delta Y = 0$ does not intersect the 95\% credibility region
(except for Model 1, where they barely do).
The 95\% credibility contour in the $\sigma_Y -\eta$ plane becomes
independent of $\sigma_Y$ for $\sigma_Y \ga 0.008$,
with 95\% credibility interval $\eta \simeq (3-6.5)\times
10^{-10}$ (allowing both for the uncertainty in the
astration of $^7$Li and in the chemical evolution of D and $^3$He).

The 95\% credibility contours in the
$\Delta Y - \eta$ plane suggest that the primeval
$^4$He abundance has been systematically underestimated, by an
amount $\Delta Y \approx +0.01$.  (Though it should be noted
that Model 1 and the lower $^7$Li abundance are just consistent
with $\Delta Y = 0$ at 95\% credibility.)  Put another way, D, $^3$He, and
$^7$Li are concordant and $^4$He is the outlayer.  (This
can also be seen in Fig.~2.)
When the likelihood function is marginalized with respect to
$\Delta Y$, the 95\% credibility
interval is $\eta \simeq (2- 6.5)\times 10^{-10}$
(allowing again for the uncertainty both in astration of $^7$Li
and in the chemical evolution of D and $^3$He).

To conclude, the fact that systematic uncertainties dominate
precludes crisp statistical statements.  The lack of a viable alternative to
the standard theory of nucleosynthesis complicates matters further
as the most powerful statistical techniques assess relative viability.
However, the rigorous techniques that we have applied
point to several conclusions that are insensitive to assumptions
made about systematic uncertainty:

\begin{itemize}

\item The predictions of the standard theory of
primordial nucleosynthesis are only consistent
with the extant observations
with 95\% credibility provided $\eta \simeq (2 - 6.5) \times 10^{-10}$.

\item Our analysis suggests that the primordial $^4$He abundance
has been systematically underestimated ($\Delta Y \approx +0.01$) or
that the random errors have been underestimated ($\sigma_Y \approx 0.01$).
Only for Model 1 (extreme destruction of $^3$He) are $\Delta Y =0$
and $\sigma_Y =0.003$ in the 95\% credibility region (cf., Figs.~5 and 6).

\item  The limit to $N_\nu$ depends upon the systematic uncertainties
in the $^4$He abundance (cf., Table~1); taking
$|\Delta Y| \le 0.02$, which is four times the estimated systematic
error and also encompasses the 95\% likelihood contour in the
$\Delta Y - \eta$ plane, leads to the 95\% credible limit $N_\nu < 3.9$.

\end{itemize}

This more rigorous analysis provides additional support for the
conclusions reached previously about the concordance interval
for $\eta$ \cite{cst}.  The limit to the number of neutrino
species is less stringent than previously quoted bounds \cite{cst,Nnu}
because we allowed a chemical evolution model with the most
extreme destruction of $^3$He (which permits low values
of $\eta$ where $^4$He production is lower) as well as a large
systematic offset in the $^4$He abundance.

Finally, there are two measurements that should
reduce the systematic uncertainties significantly,
permitting a sharper test of big-bang nucleosynthesis.  The first is
a determination of the primeval D abundance by measuring D-Ly$\alpha$
absorption due to high-redshift hydrogen clouds.  The second is
a determination of the primeval $^7$Li abundance by studying
short period, tidally locked pop II halo binaries;
astration is believed to involve rotation-driven mixing
astration and is minimized in these stars because they rotate slowly
\cite{yale}.  At the moment, there are conflicting measurements and
upper limits for the primeval D abundance seen in high-redshift
hydrogen clouds \cite{dlya}, and there is one study which
indicates that the $^7$Li abundance in short-period
binaries is no higher (evidence against significant
astration) and another that finds weak evidence that the
$^7$Li abundance is higher \cite{sptlb}.

\paragraph{Acknowledgments.}  We thank Donald Q. Lamb for many
valuable discussions and comments.  This work was supported by
the DoE (at Chicago and Fermilab) and by the NASA (at Fermilab
by grant NAG 5-2788 and at Chicago by a GSRP (CJC)).

\newpage

\section*{Figure Captions}

\bigskip
\noindent{\bf Figure 1:}  The predicted light-element abundances
(with $2\sigma$ theoretical errors); rectangles indicate
consistency intervals, which all overlap for $\eta =(2.5-6)\times 10^{-10}$.

\medskip
\noindent{\bf Figure 2:}  Likelihood functions for
D and $^3$He (lower solid curves, from left to right: Models 1, 0, and 2),
$^4$He (dotted curves, from left to right: $\sigma_Y=0.01$, $\sigma_Y=0.003$,
and $\Delta Y= 0.01$), and $^7$Li (broken = high $^7$Li, solid = low $^7$Li).

\medskip
\noindent{\bf Figure 3:}  Reduced $\chi^2$ as function of $\eta$ for
eight different sets of assumptions about the systematic uncertainties (see
text for details).

\medskip
\noindent{\bf Figure 4:}  The 5\% of maximum likelihood contours
for ${\cal L}(N_\nu, \Delta Y, \sigma_Y=0.003)$ (solid curves = low $^7$Li,
broken curves = high $^7$Li).  Because $N_\nu$ and $\Delta Y$ are not
independent parameters, the contours of likelihood are diagonal lines and
the likelihood function is not compact.

\medskip
\noindent{\bf Figure 5:}  The likelihood function ${\cal L}(\sigma_Y,
\eta ,\Delta Y=0)$ (solid curves = low $^7$Li, broken curves = high $^7$Li).

\medskip
\noindent{\bf Figure 6:}  Same as Fig.~5 for ${\cal L}(\Delta Y, \eta ,
\sigma_Y = 0.003)$.

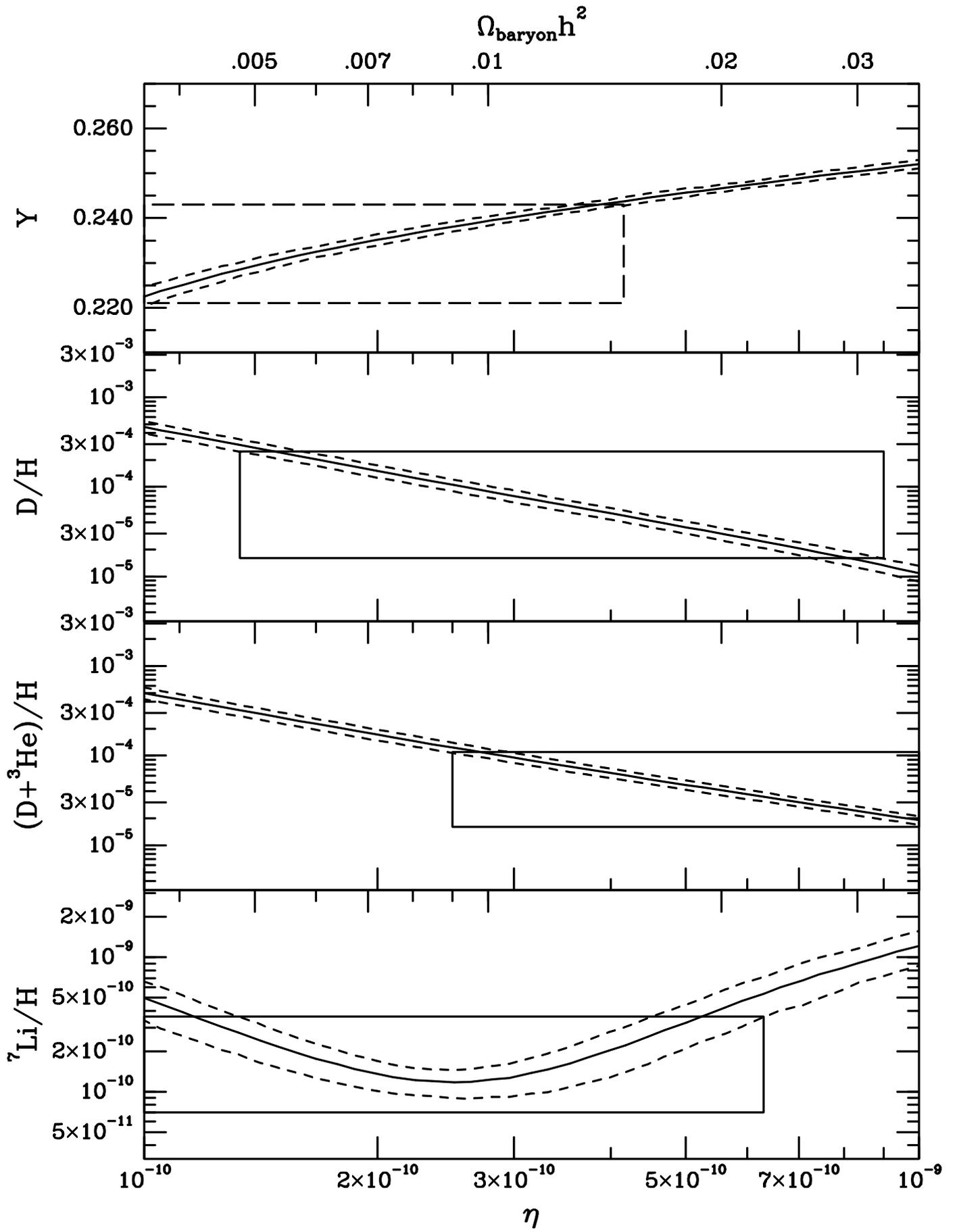
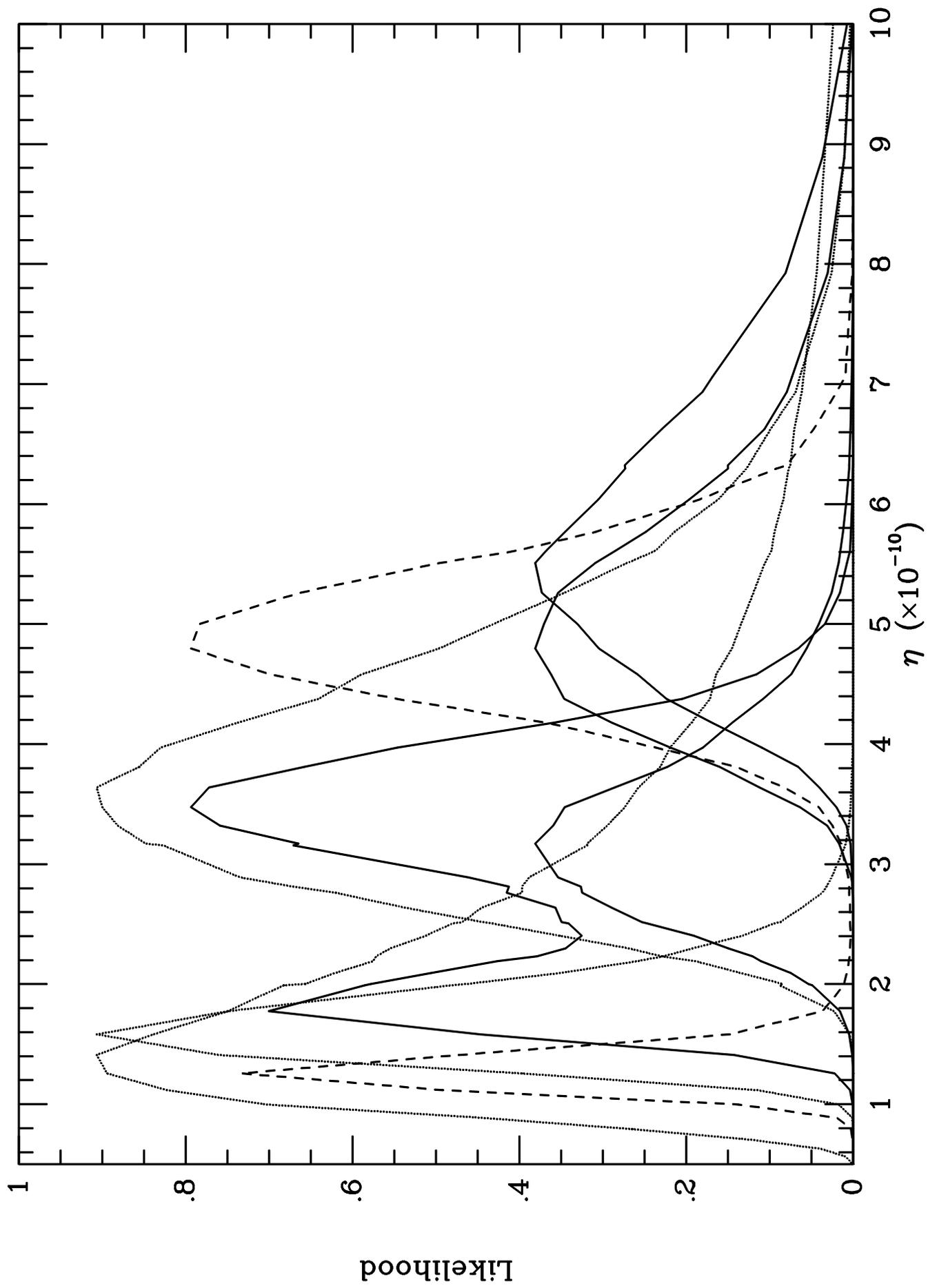
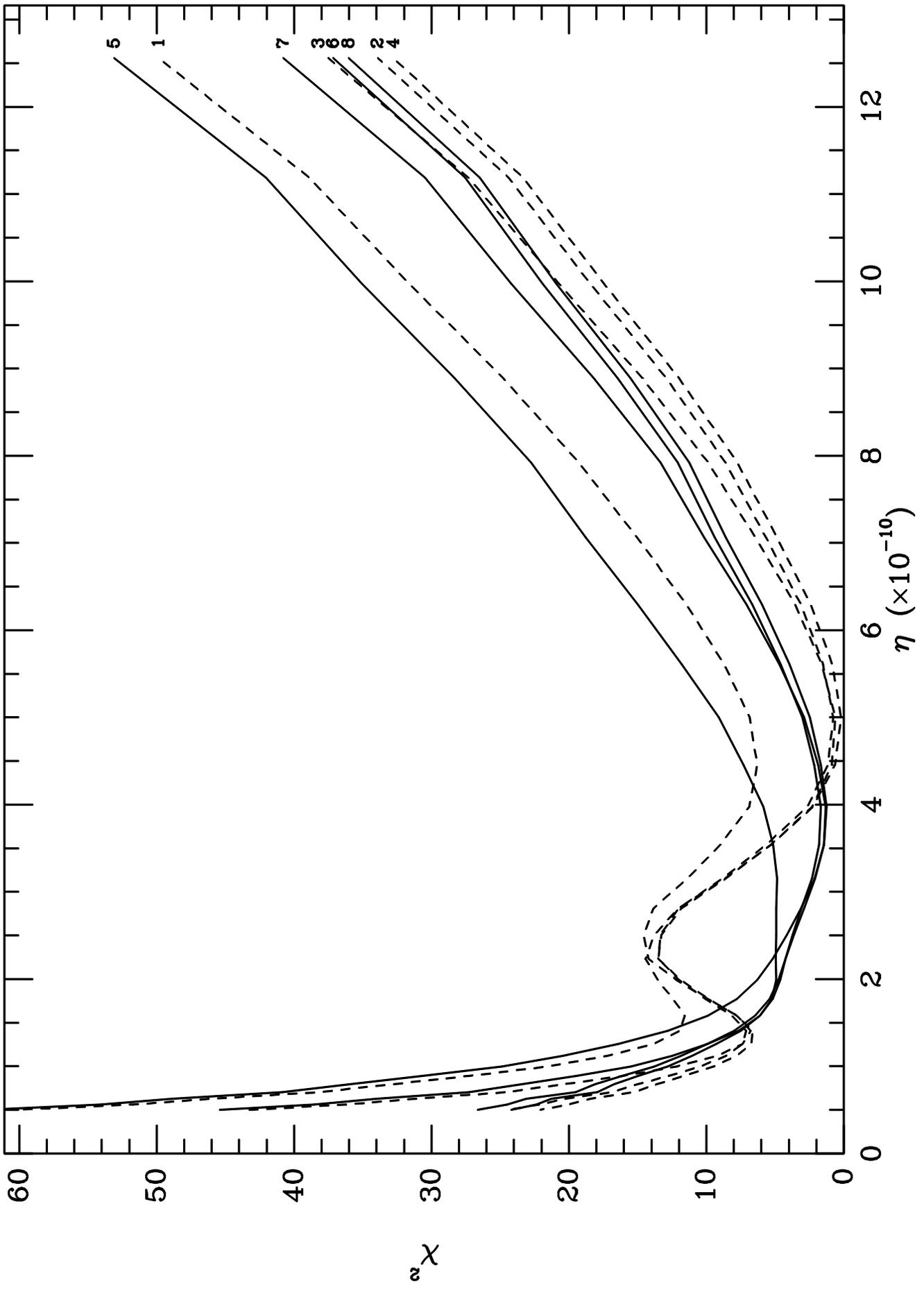
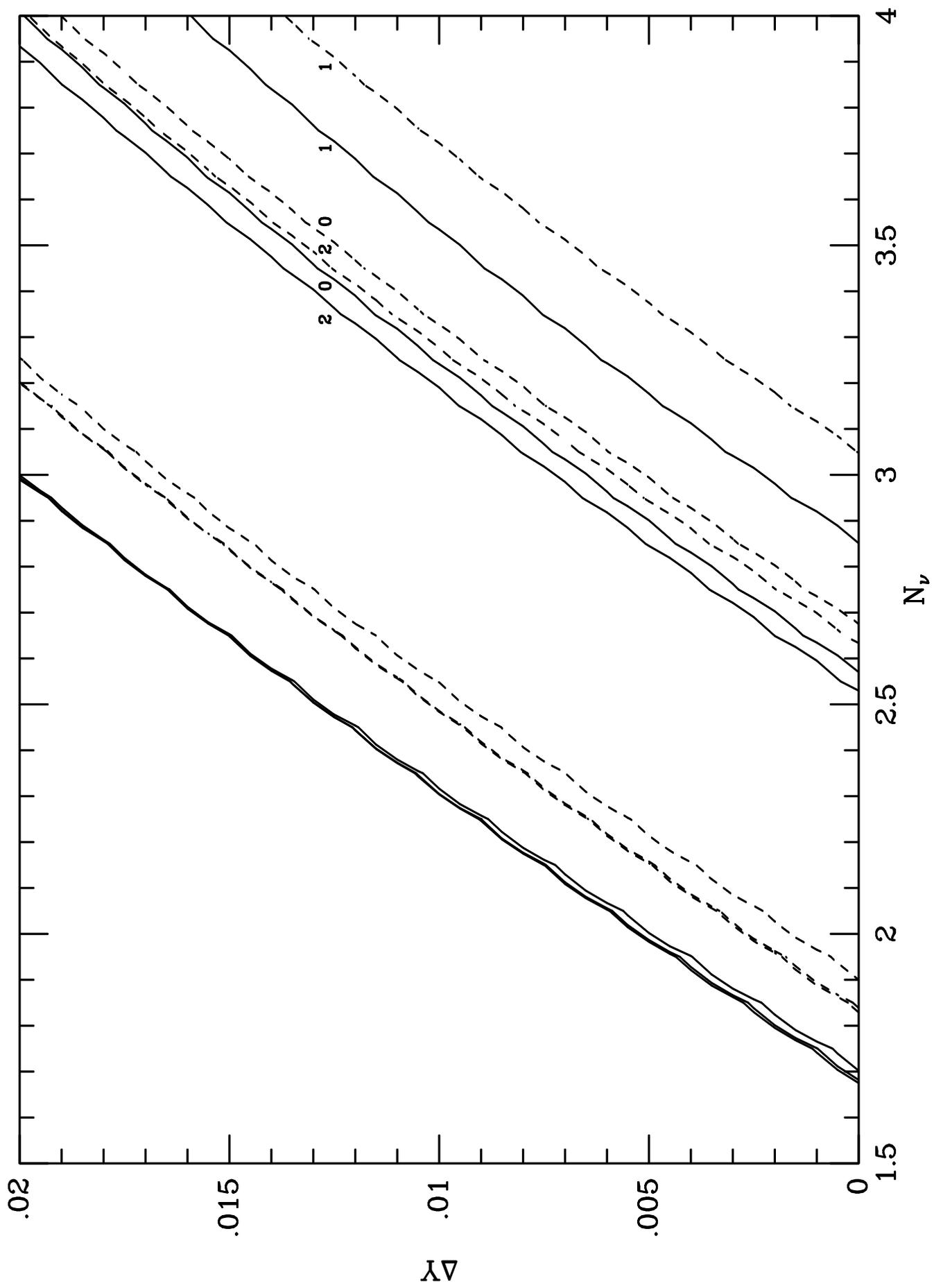
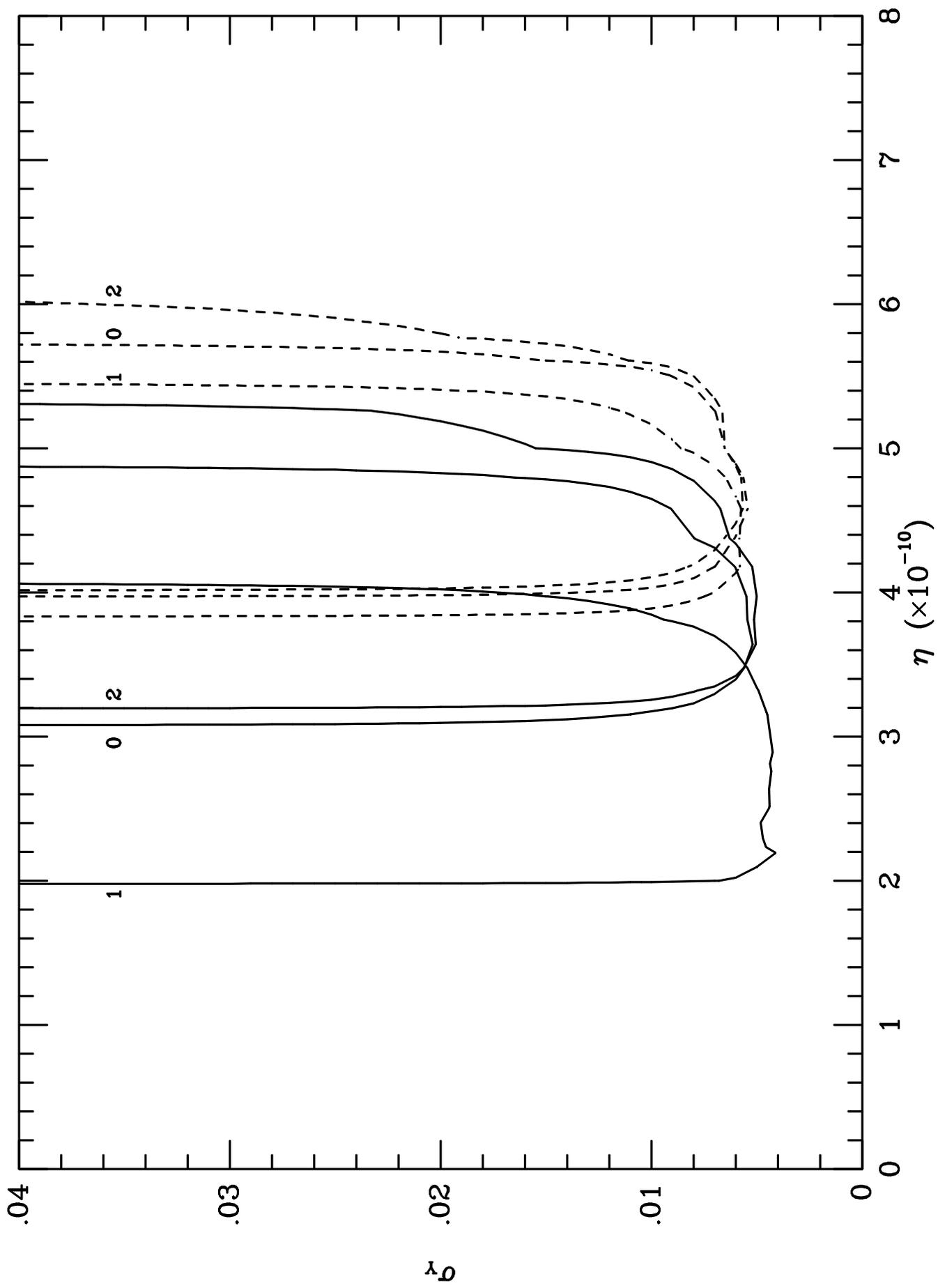
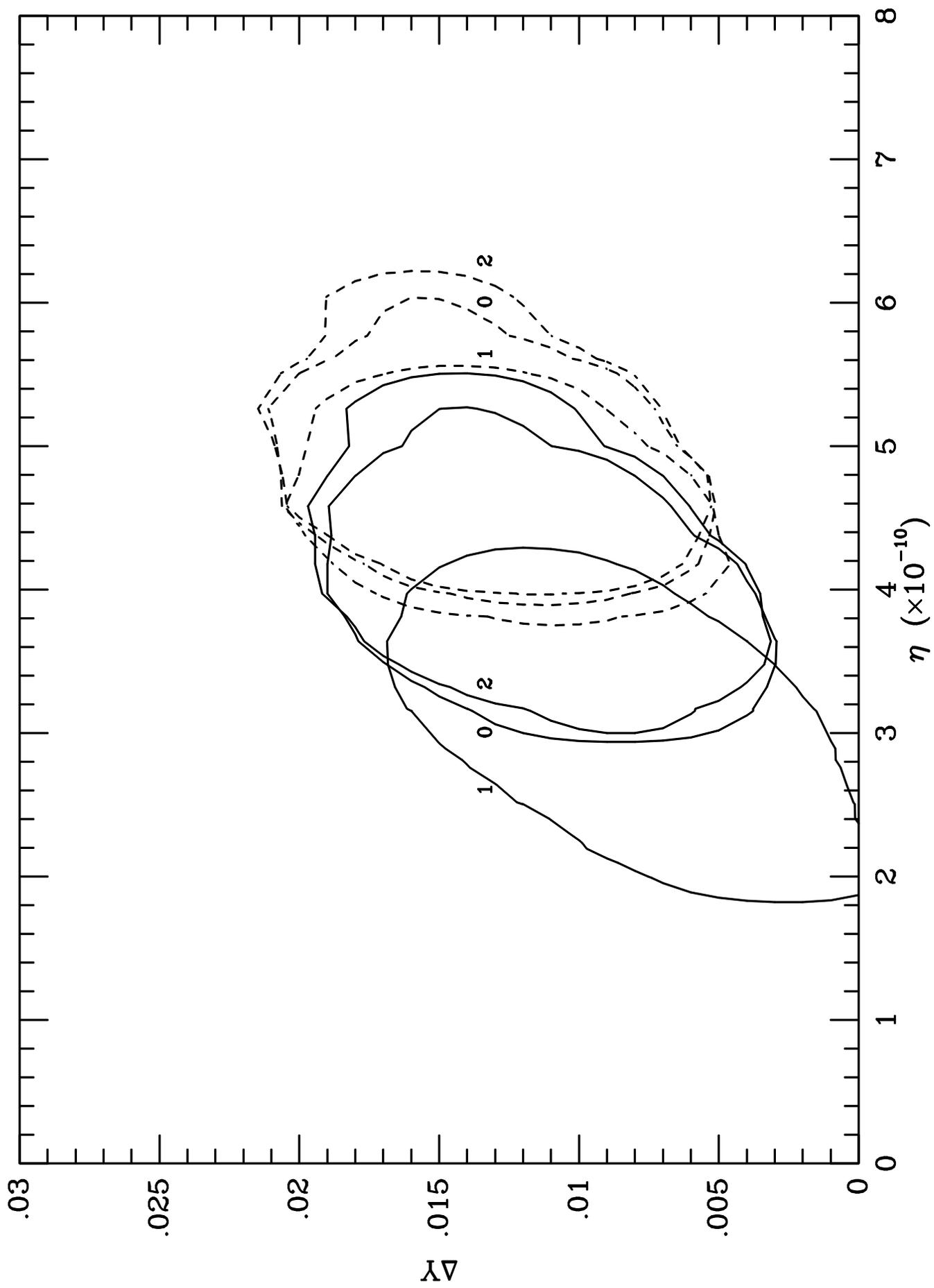
\end{document}